\pdfoutput=1

\documentclass[%
reprint,
 preprintnumbers,
 amsmath,amssymb,
 aps,
]{revtex4-2}

\usepackage{graphicx}
\usepackage{dcolumn}
\usepackage{bm}

\usepackage{amsmath,amssymb,amsbsy,amstext,amsthm,simplewick,amsfonts,graphicx}
\usepackage{mathrsfs}
\usepackage{graphicx}
\usepackage{wrapfig}
\usepackage{upgreek}
\usepackage{bm} 
\usepackage{framed}
\usepackage{bbm}
\usepackage{textcomp}
\usepackage{adjustbox}
\usepackage{makecell}
\usepackage{tcolorbox}
\usepackage{empheq}
\usepackage[normalem]{ulem}
\usepackage{enumitem}
\usepackage{braket} 
\usepackage{array}
\usepackage{dsfont}
\usepackage{physics}
\usepackage{ulem}
\usepackage{xcolor}
\usepackage{caption}
\usepackage{subcaption}
\usepackage{tocloft}
\usepackage{tikz}
\usepackage{xcolor}
\usetikzlibrary{patterns}

\usepackage{hyperref}
\hypersetup{colorlinks=true,
	breaklinks=true,
	pdfstartview=Fit,
	linkcolor=blue,
    citecolor=magenta,
	urlcolor=blue
}

\preprint{YITP-25-194}

\begin{document}
\rightline{YITP-25-194}

\title{Unveiling Primordial Black Hole Relics Through Induced Gravitational Waves}

\author{Misao Sasaki$^{1,2,3,4}$}
\email{misao.sasaki@ipmu.jp}

\author{Jianing Wang$^{1}$}
\email{jianing.wang@ipmu.jp}

\affiliation{
 $^1$ Kavli Institute for the Physics and Mathematics of the Universe (WPI), The University of Tokyo Institutes for Advanced Study, The University of Tokyo, Chiba 277-8583, Japan\\
 $^2$ Center for Gravitational Physics and Quantum Information,
 Yukawa Institute for Theoretical Physics, Kyoto University, Kyoto 606-8502, Japan\\
 $^3$ Leung Center for Cosmology and Particle Astrophysics, National Taiwan University, Taipei 10617, Taiwan\\
 $^4$ Asia Pacific Center for Theoretical Physics, Pohang 37673, Korea 
}

\date{\today}

\begin{abstract}
Black hole relics are of significant interest in cosmology and theoretical physics. In this work, we consider tiny primordial black holes (PBHs) ($M_{\mathrm{PBH}} \lesssim 10^7\, \mathrm{g}$) which are generated soon after the end of inflation and evaporate and reheat the Universe before big bang nucleosynthesis (BBN), but leave their remnants due to incomplete evaporation. 
These PBHs remnants may contribute as part or all of the dark matter (DM) today. 
Assuming that there exist PBH relics, we point out that the number density of PBH today can be directly read from the peak positions of the induced gravitational waves due to the inhomogeneous PBH distribution.
If PBH relics are of Planck mass and they forms all the DM today, the PBH number density would be of $10^{-25}\,\mathrm{cm^{-3}}$ with the peak frequency $60\,{\rm Hz}$. 
The peak frequency scales as $f_{\rm relic}^{1/3}$ where $f_{\rm relic}$ is the fraction of the PBH relics in the total DM density.
The peak amplitude carries the information of initial PBH abundance. For monochromatic-mass PBH with the current number density $10^{-41} \sim 10^{-25}\, \mathrm{cm}^{-3}$ and initial abundance $10^{-13}\sim 10^{-7}$, the amplitude may be large enough to be detected by planned gravitational wave experiments in the near future.

\end{abstract}

\maketitle


\section{Introduction}

Black hole relics are hypothetical remnants that may remain after the evaporation of a black hole via Hawking radiation. According to classical black hole thermodynamics and Hawking's seminal work \cite{Hawking:1974rv,Hawking:1975vcx,Hawking:1976de,Lowe:1992ed}, black holes are not entirely black but emit radiation due to quantum effects near the event horizon, leading to gradual mass loss and eventual evaporation. However, the final stage of this evaporation process remains uncertain \cite{Page:1976df,Page:1976ki,Page:1977um,Strominger:1996sh,Bonanno:2000ep,Bonanno:2001xi,Bonanno:2006eu,dvali2018microscopicmodelholographysurvival,Dvali:2024hsb}. One compelling idea is that instead of evaporating completely, black holes might leave behind stable or metastable remnants known as black hole relics or Planck-mass remnants \cite{Giddings:1992hh,Giddings:1993km,Rovelli:2018okm,Iozzo:2021vnq}.

These relics are of significant interest in theoretical physics because they could provide insight into the unification of quantum mechanics and general relativity, potentially serving as probes of quantum gravity effects \cite{Adler:2001vs,Wen:2015kwa,Kuntz:2019gka,Zhang:2020qxw,Zhang:2021wex,Saueressig:2021pzy,Davies:2024ysj}. One of the biggest puzzles in theoretical physics is the black hole information paradox: How information about matter that falls into a black hole is preserved or lost after it evaporates via Hawking radiation \cite{Hawking:1976ra,Banks:1992is}. 
If black holes evaporate completely, it seems information is lost \cite{Preskill:1992tc}, violating unitarity of quantum theory.
Black hole relics provide a way out: If evaporation stops (or slows down) at some stage, the remnant could store the information, preserve unitarity and resolve the paradox.
Theoretical models predicting black hole relics often incorporate modifications to standard physics, such as generalized uncertainty principles (GUP) or quantum gravity inspired corrections \cite{Scardigli:1999bt}.
In these frameworks, black hole relics are thought to form at the end stages of evaporation when quantum gravity effects dominate.
Also, it has been recently argued that the black hole evaporation suffers from the memory burden effect which slows down the evaporation rate significantly \cite{dvali2018microscopicmodelholographysurvival,Dvali:2018ytn,Dvali:2020wft,Alexandre:2024nuo,Dvali:2024hsb,Balaji:2024hpu,Haque:2024eyh,Kohri:2024qpd,Zantedeschi:2024ram,Bhaumik:2024qzd,dvali2025transitioningmemoryburdendetectable,Dondarini:2025ktz}. In this case the black hole lifetime can become much larger than the age of the Universe, hence it effectively leaves PBH relics which are much heavier than the Planck mass today.
Studying relics may offer clues about the nature of quantum gravity and the unification of general relativity with quantum mechanics-a major unsolved problem \cite{dvali2018microscopicmodelholographysurvival,Scardigli:2010gm}.

Black hole relics are also of great importance in cosmology. Primordial black holes (PBHs) could have formed in the early Universe, by various mechanisms with masses spanning a wide range \cite{Carr:1974nx,Carr:1975qj,Yokoyama:1998qw,Carr:2009jm,Carr:2016drx,Sasaki:2018dmp,Escriva:2019phb,Carr:2020gox,Escriva:2020tak,Escriva:2021aeh,Harada:2022xjp,Ozsoy:2023ryl,Harada:2024jxl,Pi:2024ert,Escriv__2024,Escriva:2024aeo,Escriva:2024lmm,Harada:2024trx,Yoo:2024lhp,Ye:2025wif,Suyama:2025dui}. 
Hawking radiation predicts that PBHs with initial masses below about $\sim10^{15}\,$g would have evaporated by today \cite{Sasaki:2018dmp}. 
However, if the evaporation stops at a finite mass, these PBHs leave behind black hole relics. 
The formation and evaporation of PBHs and their relics can influence the early Universe's thermal history, entropy budget, and matter content. 
Such relics may contribute to the dark matter (DM) content of the Universe \cite{Khlopov:2008qy,Eroshenko:2016yve,Carr:2020xqk,Pi:2022zxs}. 
If relics are sufficiently abundant and long-lived, they could account for part or all of DM without conflicting with other cosmological observations. 
This offers a non-particle-physics, gravity-based candidate for dark matter.
Even if they constitutes a fraction of DM, depending on their abundance, they could also leave imprints on cosmological observables like the cosmic microwave background, structure formation, and gravitational wave backgrounds \cite{Inman:2019wvr,Scardigli:2022jtt,Dvali:2024hsb,Domenech:2024rks,Kohri:2024qpd}.

In this work, we consider a scenario where PBHs are formed during radiation dominated era after inflation, with their lifetime much shorter than the time of big bang nucleosynthesis (BBN). 
We assume the mass spectrum to be monochromatic for simplicity.
If PBHs are abundantly produced, they eventually dominate the Universe before evaporation \cite{Carr:2017edp,Domenech:2020ssp,Perez-Gonzalez:2020vnz,Inomata:2020lmk,Arbey:2021ysg,Bernal:2021bbv,Chaudhuri:2023aiv,Conzinu:2023fui,Holst:2024ubt,Holst:2024ubt,inomata2025poltergeistmechanismenhancement}. 
We find that, if PBH remnants exist, one can extract information of the initial PBH abundance and the current number density from the peak of the induced GWs due to the initial inhomogeneous PBH distribution.

The paper is organized as follows. We briefly review the PBH formation and evaporation in Sec.~\ref{s:Formation and Evaporation}. 
We introduce PBH remnants and discuss their abundance in Sec.\ref{s:remnants}. 
Ee obtain the relation between the initial PBH mass and the relic number density. 
We discuss the characteristics of the induced GW spectrum 
and present our main result in Sec.\ref{s:IGW}. 
Section~\ref{s:Summary} is devoted to conclusions.

\section{Primordial Black Hole Formation and Evaporation}\label{s:Formation and Evaporation}

PBHs are generated in radiation dominated era shortly after the inflation. 
The initial mass of the PBHs at formation time is \cite{Sasaki_2018}
\begin{equation}\label{eq:MfHf}
M_{\mathrm{PBH}, \mathrm{f}} = \gamma \frac{4\pi}{3}\left(H_{\mathrm{f}}^{-1}\right)^3 3M_{\mathrm{pl}}^2 H_{\mathrm{f}}^{2} 
\approx 1.1 \times 10^{-5} \mathrm{~g} \frac{M_{\mathrm{pl}}}{H_{\mathrm{f}}}
\end{equation}
where $H_{\mathrm{f}}$ is the Hubble at formation time. These group of PBHs were formed around $t_{\mathrm{f}} \approx 1 / (2 H_{\mathrm{f}})$.
Reduced Planck mass: $M_{\mathrm{pl}} \equiv 1/\sqrt{8\pi G}\approx 2.435 \times 10^{18} \mathrm{GeV} \approx 4.341 \times 10^{-6} \mathrm{~g}$ \cite{Planck:2018vyg}. $\gamma\approx 0.2$ is the ratio of mass finally fell into PBH and horizon mass \cite{Carr:2020gox}. 
For extremely small PBHs which form around the end of inflation, one approximately has 
\begin{equation}
\frac{H_{\mathrm{f}}}{M_{\mathrm{pl}}} \lesssim 10^{-5} ,
\end{equation}
which implies 
\begin{equation}
M_{\mathrm{PBH}, \mathrm{f}} \gtrsim 1.1\mathrm{~g}\sim 2.5\times 10^5 M_{\mathrm{Pl}}.
\end{equation}
Define the initial energy density fraction of PBH \cite{Domenech:2023mqk}
\begin{equation}\label{def:beta}
\beta\equiv\frac{\rho_{\mathrm{PBH}, \mathrm{f}}}{\rho_{\text {tot,f}}}\,.
\end{equation}

In the conventional scenario, the early Universe is radiation dominated (energy density $\rho_{\mathrm{r}} \propto a^{-4}$, where $a$ is the scale factor) when PBHs are formed. Since PBHs behave like non-relativistic matter, with energy density scaling as $\rho_{\mathrm{PBH}} \propto a^{-3}$, even if the initial fraction of PBHs in the total energy density is small, their relative contribution grows as the Universe expands.
If PBHs are abundant enough, they can easily dominate the Universe and realize a temporary matter-dominated phase during the radiation era.
We call it an early matter-dominated era, or the PBH-dominated epoch \cite{Domenech:2020ssp}. 
The PBH-dominated epoch ends when the energy density of the radiation generated by Hawking radiation exceeds the PBH mass density, and the Universe returns to the radiation-dominated stage.

The Hawking evaporation rate is proportional to $M_{\mathrm{PBH}}^{-2}$ \cite{Hawking:1975vcx},
\begin{equation}\label{def:HawkingMt}
\frac{\mathrm{d} M_{\mathrm{PBH}}}{\mathrm{~d} t}=-\frac{A}{M_{\mathrm{PBH}}^2}, \quad A\approx 2.7~
M_{\mathrm{Pl}}^4\,.
\end{equation}
Hence we have
\begin{equation}\label{eq:HawkingMt}
M_{\mathrm{PBH}}(t)=M_{\mathrm{PBH}, \mathrm{f}}\left(1-\frac{t}{t_{\mathrm{eva}}} \right)^{1 / 3}\,,
\end{equation}
where
\begin{equation}\label{eq:teva}
t_{\mathrm{eva}} = \frac{M_{\mathrm{PBH}, \mathrm{f}}^3}{3 A}
\approx 4.1\times 10^{-28}\mathrm{~s}\left(\frac{M_{\mathrm{PBH}, \mathrm{f}}}{1 \mathrm{~g}}\right)^3.
\end{equation}
Requiring that the evaporation ends before BBN which occurs at $t_{\mathrm{eva}}\sim 1\mathrm{~s}$, the PBH mass should be smaller than $\sim 10^9 \mathrm{g}$.

\begin{figure}
\includegraphics[width=9cm]{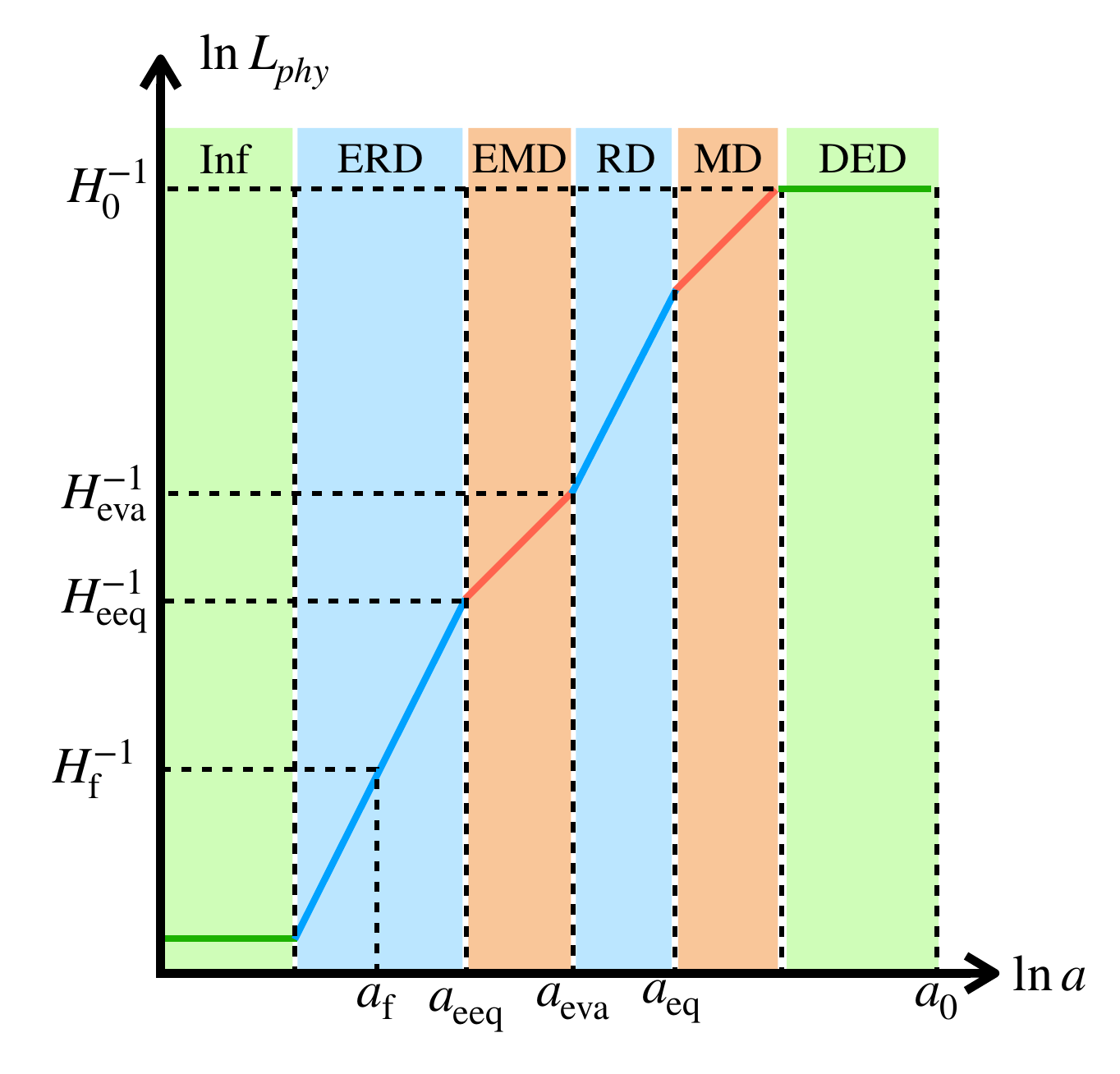}
\caption{A schematic diagram of PBH-dominated era, or early matter dominated era (EMD), when $H_{\mathrm{eva}}\to H_{\mathrm{eeq}}$, the PBH-dominated era duration would become shorter and shorter. Inf: inflation, ERD: early radiation dominated era, RD: radiation dominated era, MD: matter dominated era, DED: dark energy dominated era, eeq: early rad-matter equilibrium, eq: rad-matter equilibrium. } 
\label{fig:schematic}
\end{figure}

One important variable is the ratio of the PBH lifetime to the formation time, 
\begin{equation}\label{def:Revaf}
\frac{t_{\mathrm{eva}}}{t_{\mathrm{f}}}
\approx3.3 \times 10^{10} \left(\frac{M_{\mathrm{PBH}, \mathrm{f}}}{1 \mathrm{~g}}\right)^2 \gg 1\,,
\end{equation}
which makes it easy to realize a PBH-dominated era.
If there is indeed a PBH-dominated era, we can set $a_{\mathrm{eeq}}$ as the time of early radiation-PBH equality. 
Then at early equality time and PBH formation time, we have
\begin{equation}
3 M_{\mathrm{pl}}^2 H_{\mathrm{eeq}}^2 = \rho_{\mathrm{PBH,eeq}}+\rho_{\mathrm{r,eeq}}=2 \rho_{\mathrm{PBH,eeq}},
\end{equation}
and 
\begin{equation}
\begin{aligned}\label{eq:explain12}
3 M_{\mathrm{pl}}^2 H_{\mathrm{f}}^2 =&\rho_{\mathrm{PBH,f}}+\rho_{\mathrm{r,f}}\\
=&\rho_{\mathrm{PBH,eeq}}\left[\left(\frac{a_{\mathrm{f}}}{a_{\mathrm{eeq}}}\right)^{-3}+\left(\frac{a_{\mathrm{f}}}{a_{\mathrm{eeq}}}\right)^{-4}\right]\,.
\end{aligned}
\end{equation}
Combining them and using the definition \eqref{def:beta} of $\beta$, we obtain 
\begin{equation}
\frac{1}{\beta} =1+\left(\dfrac{a_{\mathrm{f}}}{a_{\mathrm{eeq}}}\right)^{-1}.
\end{equation}
Considering $ \beta \ll 1$, we have 
\begin{equation}\label{eq:betanew}
\frac{a_{\mathrm{f}}}{a_{\mathrm{eeq}}} 
\approx \beta\,,
\end{equation}
or \cite{Domenech:2023mqk}
\begin{equation}\label{eq:beta}
\frac{H_{\mathrm{eeq}}^2}{H_{\mathrm{f}}^2} =\frac{2}{\left(\dfrac{a_{\mathrm{f}}}{a_{\mathrm{eeq}}}\right)^{-3}+\left(\dfrac{a_{\mathrm{f}}}{a_{\mathrm{eeq}}}\right)^{-4}} 
\approx 2 \beta^4.
\end{equation}
The evaporation can happen during radiation dominated era or PBH dominated era, depending on the initial energy density fraction of PBH.  We focus on the case where PBHs dominate the universe before evaporation.
It can be seen from Fig.\ref{fig:schematic}, the condition for the appearance of a PBH dominated era is 
\begin{equation}
H_{\mathrm{eva}} \leq H_{\mathrm{eeq}}.
\end{equation}
Noting that ${H_{\mathrm{f}}}/{H_{\mathrm{eva}}}=3t_{\mathrm{eva}}/4t_{\rm f}$,
the minimum PBH initial abundance is
\begin{equation}\label{def:betamin}
\beta_{\mathrm{min}} \approx \left( \frac{H_{\mathrm{eva}}}{\sqrt{2}H_{\mathrm{f}}}\right)^{1/2} \approx 5.3\times 10^{-6}\left(\frac{M_{\mathrm{PBH}, \mathrm{f}}}{1 \mathrm{~g}}\right)^{-1}\,.
\end{equation}

\begin{figure}
\includegraphics[width=9cm]{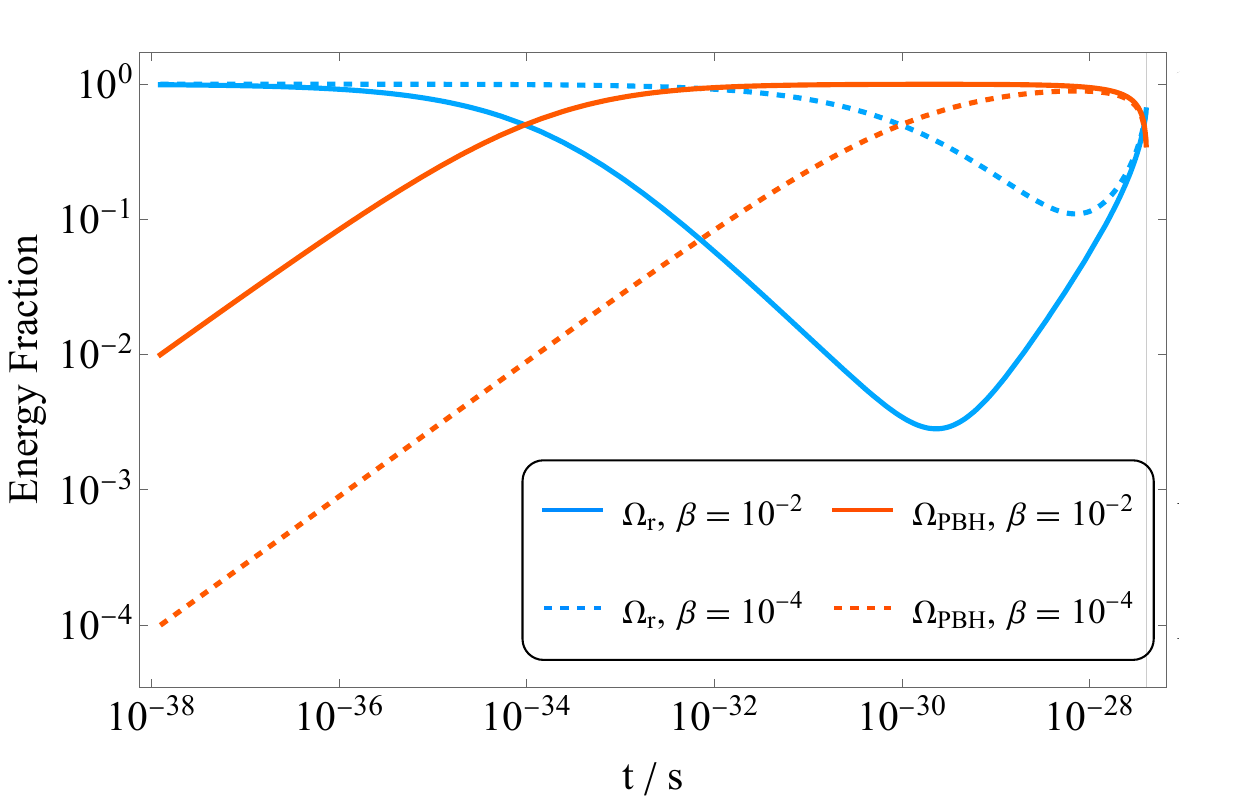}
\caption{Energy fraction in two examples described by \eqref{eq:drhoPBH} and \eqref{eq:drhor}. In all lines, $M_{\mathrm{PBH,f}}=1\mathrm{~g}$, starting from $t_{\mathrm{f}}$, ending with $t_{\mathrm{eva}}$ (vertical line). The blue lines are the time evolution of energy fraction $\Omega_{\mathrm{r}}$, the orange ones are for $\Omega_{\mathrm{PBH}}$. The solid lines is an example for $\beta=10^{-2}$, it appears a PBH-dominated era. The dashed lines is an example for $\beta=10^{-4}$ \eqref{def:betamin}, PBH-dominated era is shorter.} 
\label{fig:energyfraction}
\end{figure}

Now we consider the energy density evolution.
Let $n_{\mathrm{PBH}}$ be the PBH number density. Since the PBH energy density is
given by $\rho_{\mathrm{PBH}}=n_{\mathrm{PBH}} M_{\mathrm{PBH}}$,
we have
\begin{equation}\label{eq:drhoPBH}
\frac{\mathrm{d} \rho_{\mathrm{PBH}}}{\mathrm{d} t}=-3 H \rho_{\mathrm{PBH}}+n_{\mathrm{PBH}} \frac{\mathrm{d} M_{\mathrm{PBH}}}{\mathrm{d} t}\,,
\end{equation}
where the second term represents the effect of the evaporation.
In turn, the radiation gains energy from the PBH evaporation, which balances the second term in \eqref{eq:drhoPBH} to give
\begin{equation}\label{eq:drhor}
\frac{\mathrm{d} \rho_{\mathrm{r}}}{\mathrm{d} t}=-4 H \rho_{\mathrm{r}}-n_{\mathrm{PBH}} \frac{\mathrm{~d} M_{\mathrm{PBH}}}{\mathrm{~d} t}.
\end{equation}
Solving \eqref{eq:drhoPBH} and \eqref{eq:drhor} \cite{Hooper:2019gtx,Dom_nech_2021}, one can obtain the energy density evolution from the PBH formation time to evaporation time with specific choice of $\beta$ value, see Fig.\ref{fig:energyfraction}.

\section{PBH relics as dark matter}\label{s:remnants}

We assume that the PBH evaporation leaves behind relics with mass given by
\begin{equation}
m_{\text {relic}}=r M_{\mathrm{pl}}\,; \quad r\geq 1.
\end{equation}
Introducing the time when the evaporation stops,
\begin{equation}\label{def:mrelic}
M_{\mathrm{PBH}}\left(t_{\text {end}}\right)=m_{\text {relic}}\,,
\end{equation}
we find the difference between the time $t_{\rm eva}$ and $t_{\rm relic}$ is small,
\begin{equation} \label{eq:deltat}
\frac{\delta t}{t_{\rm eva}} =1-\frac{t_{\mathrm{end}}}{t_{\rm eva}}
=\frac{m_{\rm relic}^3}{M_{{\rm PBH,f}}^3}\ll1\,,
\end{equation}
for $m_{\rm relic}\ll M_{{\rm PBH,f}}$.

Now we study the abundance of relics.
Let $f_{\text {relic}}$ be the fraction of the PBH relics in the total DM density today,
\begin{equation}\label{def:frelic}
f_{\text {relic}} \equiv \frac{\rho_{\text {relic,0}}}{\rho_{\mathrm{DM,0}}}\,,
\end{equation}
where the subscript $0$ denotes the present value.
Following \cite{Domenech:2023mqk}, we derive the relation between $M_{{\rm PBH},f}$ and $f_{\text {relic}}$. 
The key quantity is the fraction of the relic density at the moment of evaporation, 
\begin{equation}\label{def:Omegareliceva}
\left.\Omega_{\text {relic}}\right|_{\text {eva }}\equiv\frac{\rho_{\text {relic,eva}}}{\rho_{\text {tot,eva}}}\,.
\end{equation}
The idea is to express this both in terms of $\Omega_{\rm relic,0}=f_{\rm relic}\Omega_{\rm DM,0}$ and in terms of $\Omega_{{\rm PBH},f}=\beta$, and identify them.

First, we consider the evolution from the time of evaporation to today. Since $\rho_{\text {relic}} \propto a^{-3}$, one has 
\begin{equation}\label{eq:Omegareliceva}
\left.\Omega_{\text {relic}}\right|_{\text {eva}}=f_{\text {relic }}\Omega_{\rm DM,0}= \frac{\rho_{\mathrm{DM}}\left(T_0\right)}{\rho_{\text {tot}}\left(T_{\text {eva}}\right)}\left(\frac{a_0}{a_{\text{eva }}}\right)^3,
\end{equation}
where the scale factor ratio $a_0/a_{\rm eva}$ can be expressed in terms of the temperature ratio as
\begin{equation}\label{eq:entropycons}
\frac{a_{\mathrm{eva}}}{a_0}=\left(\frac{g_{s \star}\left(T_0\right)}{g_{s \star}\left(T_{\mathrm{eva}}\right)}\right)^{1 / 3} \frac{T_0}{T_{\mathrm{eva}}}\,.
\end{equation}
Together with  
\begin{equation}
\rho_{\text {tot}}\left(T_{\text {eva}}\right)=\frac{\pi^2}{30} g_{\star}\left(T_{\text {eva}}\right) T_{\text {eva}}^4\,,
\end{equation}
one obtains
\begin{equation}
\begin{aligned}\label{eq:Omegarelic1}
&\left.\Omega_{\text {relic}}\right|_{\text {eva}} 
\approx 5.8 \times 10^{-13} f_{\text {relic }}\left(\frac{T_{\text {eva}}}{1 \mathrm{TeV}}\right)^{-1}
\\
&\qquad\times\left(\frac{g_{s \star}\left(T_{\text {eva}}\right)}{106.75}\right)\left(\frac{g_{\star}\left(T_{\text {eva}}\right)}{106.75}\right)^{-1}\,,
\end{aligned}
\end{equation}
where we adopted the values, $\Omega_{\mathrm{DM}, 0} h^2=0.120$ \cite{Planck:2018vyg}, $T_0=1.2644 \times 10^{-6} \mathrm{GeV}$, and $g_{s \star}\left(T_0\right)=3.91$.

The temperature right after the moment of PBH evaporation $T_{\mathrm{eva}}$ can be estimated by using
\begin{equation}\label{eq:energycons}
3 M_{\mathrm{Pl}}^2 H_{\mathrm{eva}}^2=\rho_{\rm tot}(T_{\rm eva})=\frac{\pi^2 g_{\star}(T_{\mathrm{eva}})}{30} T_{\mathrm{eva}}^4\,.
\end{equation}
This gives
\begin{equation}\label{eq:Teva}
T_{\mathrm{eva}} \approx 2.8 \times 10^{10} \mathrm{GeV} \left(\frac{M_{\mathrm{PBH}, \mathrm{f}}}{1 \mathrm{~g}}\right)^{-3 / 2}\left(\frac{g_{ \star}\left(T_{\mathrm{eva}}\right)}{106.75}\right)^{-1 / 4}.
\end{equation}
For example, $T_{\mathrm{eva}}\sim 1\mathrm{GeV}$ for $M_{\mathrm{PBH,f}}\sim 10^7 \,\mathrm{g}$. 
Substituting the above into \eqref{eq:Omegarelic1}, one obtains
\begin{equation}
\begin{aligned}\label{eq:Omega-backward}
&\left.\Omega_{\text {relic}}\right|_{\text {eva}}\approx 2.1\times 10^{-20} f_{\text {relic}} \left(\frac{M_{\mathrm{PBH}, \mathrm{f}}}{1\, \mathrm{g}}\right)^{3 / 2}
\\
&\qquad\times\left(\frac{g_{ \star}\left(T_{\mathrm{eva}}\right)}{106.75}\right)^{-3 / 4} \left(\frac{g_{s \star}\left(T_{\text {eva}}\right)}{106.75}\right).
\end{aligned}
\end{equation}

Now we turn to the evolution from the PBH formation to evaporation.
The energy density of PBH relics at the evaporation time is
\begin{equation}
\begin{aligned}
 \rho_{\text {relic,eva}}
 &=m_{\text {relic}} n_{\mathrm{PBH}, \text{eva}}
 \\
&=m_{\text {relic}} n_{\mathrm{PBH,f}}\left(\frac{a_{\mathrm{f}}}{a_{\mathrm{eva}}}\right)^3\,,
\end{aligned}\label{eq:rho_eva}
\end{equation}
where the PBH number density at the formation time is expressed as
\begin{equation}\label{def:nPBHf}
n_{\mathrm{PBH}, \mathrm{f}}\equiv\frac{\rho_{\mathrm{PBH}, \mathrm{f}}}{M_{\mathrm{PBH}, \mathrm{f}}}
= \frac{3 \beta}{4 \pi \gamma} H_{\mathrm{f}}^3\,.
\end{equation}
From the above, the definition of $\Omega_{\rm relic}|_{\rm eva}$ \eqref{def:Omegareliceva} gives \cite{Domenech:2023mqk}
\begin{equation}\label{eq:Omega-foreward}
\left.\Omega_{\text {relic}}\right|_{\text {eva }}=\frac{\beta\, m_{\text {relic}}}{M_{\text {PBH,f}}} \frac{H_{\mathrm{f}}^2}{H_{\text {eva}}^2} \left(\frac{a_{\text {f}}}{a_{\text {eva}}}\right)^3\,.
\end{equation}
Assuming that the PBH evaporation ends after the universe becomes dominated by PBHs, using \eqref{eq:betanew} and \eqref{eq:beta} one can prove 
\begin{equation}\label{eq:aevaaf-PBH}
\beta \frac{H_{\mathrm{f}}^2}{H_{\mathrm{eva}}^2} \left(\frac{a_{\mathrm{f}}}{a_{\mathrm{eva}}}\right)^3=1\,.
\end{equation}
Thus
\eqref{eq:Omega-foreward} gives a very simple formula,
\begin{equation}\label{eq:Omega-foreward-2}
\left.\Omega_{\text {relic}}\right|_{\text {eva}}=\frac{m_{\text {relic}}}{M_{\mathrm{PBH}, \mathrm{f}}}\, .
\end{equation}

Combining \eqref{eq:Omega-backward} and \eqref{eq:Omega-foreward-2}, the initial PBH mass is uniquely determined as a function of $r/f_{\rm relic}$,
\begin{equation}
\begin{aligned}\label{eq:subMf}
&\frac{M_{\mathrm{PBH}, \mathrm{f}}}{1\, \mathrm{g}} =5.3\times 10^{5} \left(\frac{r}{f_{\text {relic}}}\right)^{2/5}
\\
&\qquad\times\left(\frac{g_{ \star}\left(T_{\mathrm{eva}}\right)}{106.75}\right)^{3 / 10} \left(\frac{g_{s \star}\left(T_{\text {eva}}\right)}{106.75}\right)^{-2/5}.
\end{aligned}
\end{equation}

\section{Relic Abundance Hint From IGW}\label{s:IGW}

A PBH-dominated era can leave a distinct imprint on the stochastic GW background. The random distribution of PBHs naturally leads to the Poisson power spectrum of the PBH density perturbation, which is isocurvature at the time of PBH formation.  It induces GWS at second order in perturbation \cite{Ananda_2007, Inomata_2019,Domenech:2020ssp,Arimoto:2021cwc}, which can be potentially detected by future GW observations.

The dimensionless PBH density perturbation spectrum at the formation moment is given by \cite{Dom_nech_2021}
\begin{equation}\label{eq:PSofS}
\mathcal{P}_S(k)\equiv\frac{k^3}{2 \pi^2} P_S(k)=\frac{k^3}{2 \pi^2} \frac{1}{n_{\mathrm{PBH,f}} a_{\mathrm{f}}^3}=\frac{2}{3 \pi}\left(\frac{k}{k_{\mathrm{uv}}}\right)^3,
\end{equation}
where 
the ultra-violet (UV) cut-off $k_{\rm uv}$ is given by the mean inter-PBH separation at the formation time,
\begin{equation}\label{eq:kuv}
k_{\mathrm{uv}} \equiv a_{\mathrm{f}}\left(\frac{4\pi}{3} n_{\mathrm{PBH,f}}\right)^{1/3} 
= \frac{\beta^{1 / 3}}{\gamma^{1 / 3}} a_{\mathrm{f}} H_{\mathrm{f}}\,.
\end{equation}
The PBH density perturbations are isocurvature in the beginning, but they evolve to curvature perturbations as PBHs begin to dominate the universe. 
When the PBH evaporation renders the universe radiation-dominated again, 
those curvature perturbations start oscillating rapidly and produce GWs. 
If we assume the complete evaporation, the induced GW spectrum today has a peak at $k_{\mathrm{uv}}$ with the peak amplitude \cite{Dom_nech_2021,Domenech:2023mqk},
\begin{equation}
\begin{aligned}\label{eq:igw0}
\Omega_{\mathrm{GW}, 0}^{\mathrm{peak}} h^2\approx& 2.1 \times 10^{-11} \beta^{16 / 3}\left(\frac{k_{\mathrm{uv}}}{k_{\mathrm{eva}}}\right)^{17 / 3}\\
\cdot& \left(\frac{g_{\star}\left(T_{\mathrm{eva}}\right)}{106.75}\right)\left(\frac{g_{s \star}\left(T_{\mathrm{eva}}\right)}{106.75}\right)^{-4 / 3},
\end{aligned}
\end{equation}
where $k_{\rm eva}$ is the comoving wavenumber that crosses the Hubble horizon at the time of evaporation, $k_{\rm eva}=H(t_{\rm eva})a(t_{\rm eva})$.

Now we consider the case where the PBH evaporation is incomplete and PBHs leave behind their remnants. 
The physical frequency today $f$ is given by the comoving wavenumber $k$
as $f=k/ (2 \pi a_0)$.
From \eqref{eq:entropycons} and \eqref{eq:energycons}, the frequency $f_{\rm eva}$ today is expressed as
\begin{equation}
f_{\mathrm{eva}}=\frac{H_{\mathrm{eva}}}{2 \pi} \left(\frac{g_{s \star}\left(T_0\right)}{g_{s \star}\left(T_{\mathrm{eva}}\right)}\right)^{1 / 3} \frac{T_0}{T_{\mathrm{eva}}} .
\end{equation}
Substitute $T_0 \approx 2.35 \times 10^{-13} \mathrm{GeV}$, $g_{s \star}\left(T_0\right) \approx 3.91$, one finally have 
\begin{equation}
\begin{aligned}
f_{\mathrm{eva}} \approx& 2.7 \times 10^{-5} \mathrm{~Hz}\left(\frac{T_{\mathrm{eva}}}{1 \mathrm{TeV}}\right)\\
\times&\left(\frac{g_{\star}\left(T_{\mathrm{eva}}\right)}{106.75}\right)^{1 / 2}\left(\frac{g_{s \star}\left(T_{\mathrm{eva}}\right)}{106.75}\right)^{-1 / 3}.
\end{aligned}
\end{equation}
Taking the ratio of \eqref{eq:kuv} and $k_{\mathrm{eva}}=a_{\mathrm{eva}} H_{\mathrm{eva}}$, we find
\begin{equation}\label{eq:kuvkeva0}
\frac{k_{\mathrm{uv}}}{k_{\mathrm{eva}}}=\left(\frac{H_{\mathrm{f}}}{\gamma H_{\mathrm{eva}}}\right)^{1 / 3}\,.
\end{equation}
where we have used the identity \eqref{eq:aevaaf-PBH}.
Using \eqref{def:Revaf}, we obtain
\begin{equation}\label{eq:kuvkeva}
\frac{k_{\mathrm{uv}}}{k_{\mathrm{eva}}} \approx 5.0\times 10^3 \left(\frac{M_{\mathrm{PBH}, \mathrm{f}}}{1 \mathrm{~g}}\right)^{2 / 3}.
\end{equation}
This gives 
\begin{equation}
\begin{aligned}\label{eq:fuv0}
f_{\mathrm{uv}} =&f_{\mathrm{eva}} \frac{k_{\mathrm{uv}}}{k_{\mathrm{eva}}}
\approx  3.7\times 10^6\mathrm{~Hz} ~\left(\frac{M_{\mathrm{PBH}, \mathrm{f}}}{1 \mathrm{~g}}\right)^{-5 / 6}\\
&\times\left(\frac{g_{\star}\left(T_{\mathrm{eva}}\right)}{106.75}\right)^{1 / 4}\left(\frac{g_{s \star}\left(T_{\mathrm{eva}}\right)}{106.75}\right)^{-1 / 3}\,.
\end{aligned}
\end{equation}
Using \eqref{eq:subMf} further, one obtains 
\begin{equation}
\label{eq:fuv}
f_{\mathrm{uv}} 
\approx  61.9 \mathrm{~Hz} ~\left(\frac{r}{f_{\text {relic }}}\right)^{-1 / 3}\,.
\end{equation}
This is one of our main findings. The peak frequency of the induced GWs depends only on the relic mass and its current abundance. 
The case $r=f_{\rm relic}=1$ was derived in \cite{Domenech:2023mqk}.

Substituting \eqref{eq:kuvkeva} into \eqref{eq:igw0}, and using \eqref{eq:subMf}, the peak amplitude of the GW spectrum is given as
\begin{equation}
\begin{aligned}\label{eq:igwoh2}
\Omega_{\mathrm{GW}, 0}^{\mathrm{peak}} h^2
&\approx 1.7\times 10^{-11} \left(\frac{\beta}{10^{-8}}\right)^{16 / 3}  \left(\frac{r}{f_{\text {relic }}}\right)^{68 / 45} \\
\times&\left(\frac{g_{\star}\left(T_{\text {eva }}\right)}{106.75}\right)^{32 / 15}\left(\frac{g_{s \star}\left(T_{\text {eva }}\right)}{106.75}\right)^{-128 / 45}.
\end{aligned}
\end{equation}
According to Fig.~5 in \cite{Saikawa_2018}, $g_{s \star} \approx g_{\star} $ is always a good approximation in the standard model above 1\,Mev. 
For $T>T_{\mathrm{EW}}\sim 200 \mathrm{GeV}$, $g_{s \star} \approx g_{\star}\approx 106.75$. 
For $T<1\, \mathrm{GeV}$, the value of $g_{s \star}$ (or $g_{\star}$) drops down by about one order of magnitude. 
In this work, in order to avoid unnecessary complications, we focus on $1 \mathrm{~g} \lesssim M_{\mathrm{PBH}, \mathrm{f}} \lesssim 10^7 \mathrm{~g}$ ($1 \lesssim  r \lesssim 10^{12}$), which corresponds to $T_{\mathrm{eva}}\gtrsim 1 \mathrm{GeV}$. 
Then we have
\begin{equation}
\begin{aligned}\label{eq:betacontour0}
&\frac{r}{f_{\text {relic }}}
\approx  \left(\frac{f_{\mathrm{uv}}}{61.9\mathrm{~Hz} }\right)^{-3}\,,\\
&~\beta \approx  10^{-8}\left(\frac{\Omega_{\mathrm{GW}, 0}^{\mathrm{peak}} h^2}{1.7\times 10^{-11} }\right)^{3 / 16} \left(\frac{f_{\mathrm{uv}}}{61.9\mathrm{~Hz} }\right)^{17 / 20} .
\end{aligned}
\end{equation}
The above are the summary of our results. The GW peak frequency determines the ratio of the relic PBH mass and its abundance today, while the amplitude and the peak frequency tells us the efficiency of the PBH formation in the early universe.

Finally, it is useful to mention that there is one-to-one correspondence between $r/f_{\text {relic}}$ and the PBH number density today,
\begin{equation}
n_{\mathrm{PBH,0}} = \frac{f_{\mathrm{relic}}}{r}\frac{\rho_{\mathrm{DM,0}}}{M_{\mathrm{Pl}}}
\approx  5.2 \times 10^{-25} \mathrm{~cm}^{-3}\frac{f_{\mathrm{relic}}}{r}.
\end{equation}
In terms of the PBH number density, \eqref{eq:fuv} and \eqref{eq:igwoh2} become
\begin{equation}
\begin{aligned}\label{eq:betacontour}
&f_{\mathrm{uv}} 
\approx   61.9 \mathrm{~Hz} ~\left(\frac{n_{\mathrm{PBH,0}}}{5.2 \times 10^{-25} \mathrm{~cm}^{-3}}\right)^{1 / 3},\\
&\Omega_{\mathrm{GW}, 0}^{\mathrm{peak}} h^2\approx 1.7\times 10^{-11} \left(\frac{\beta}{10^{-8}}\right)^{16 / 3} \\
&\quad\qquad\times\left(\frac{n_{\mathrm{PBH,0}}}{5.2 \times 10^{-25} \mathrm{~cm}^{-3}}\right)^{-68 / 45} .
\end{aligned}
\end{equation}

\begin{figure}
\includegraphics[width=7cm]{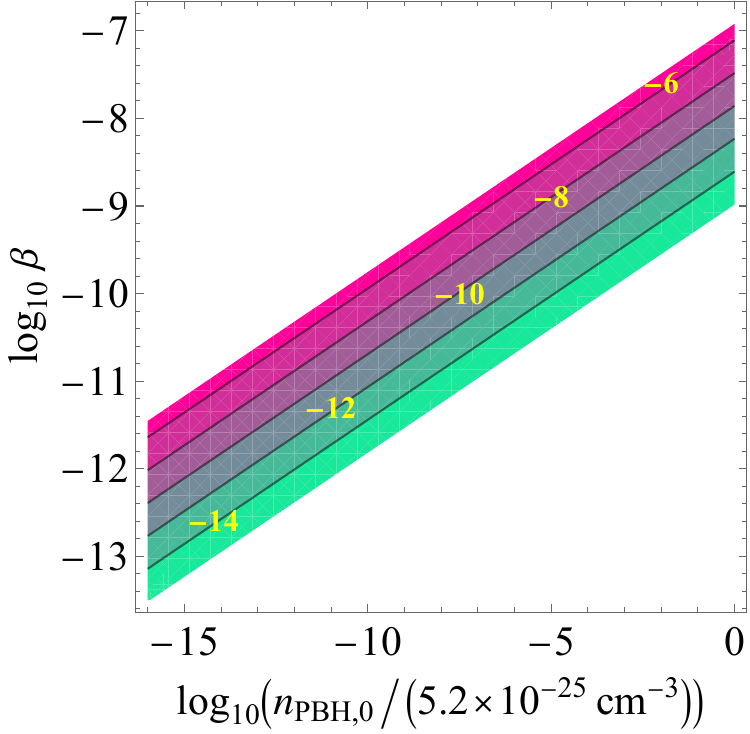}
\caption{Contour plot of $\Omega_{\mathrm{GW}, 0}^{\text {peak }} h^2$ in the $(\beta,n_{\mathrm{PBH}, 0})$ space. 
We focused on the region $10^{-16} < \Omega_{\mathrm{GW}, 0}^{\text {peak }} h^2 < 10^{-5}$. }
\label{fig:contour-beta-rfrelics}
\end{figure}
\begin{figure}
\includegraphics[width=7cm]{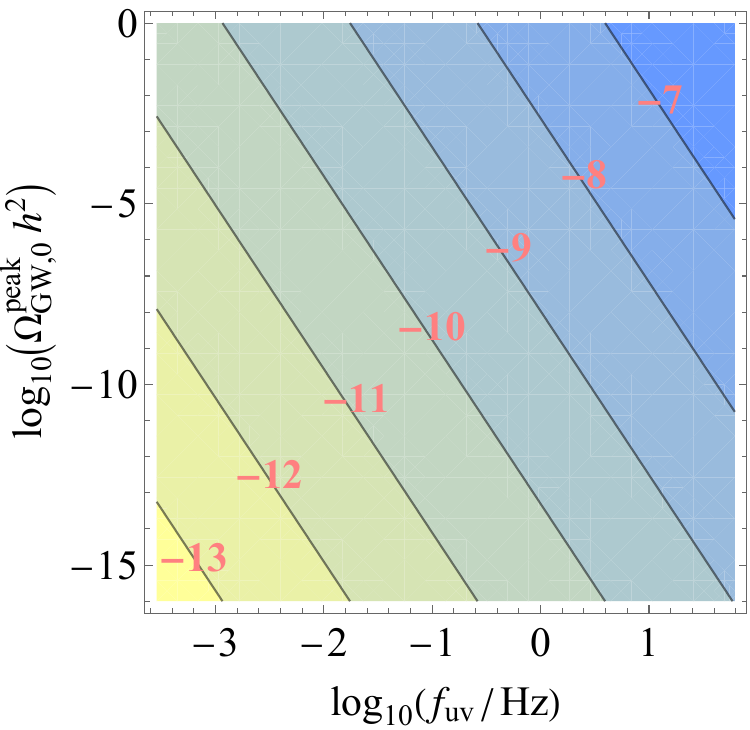}
\caption{
Contour plot of $\beta$ in the $(f_{\text {uv}},\Omega_{\mathrm{GW}, 0}^{\text {peak }} h^2)$ plane.
} 
\label{fig:contour-Omega-fuv}
\end{figure}

We show $\Omega_{\text {GW,} 0}^{\text {peak}} h^2$ as a function of $(\beta,n_{\mathrm{PBH}, 0})$ or $(\beta,f_{\mathrm{relic}}/r)$ in Figs.~\ref{fig:contour-beta-rfrelics} and \ref{fig:contour-Omega-fuv}. 
The relevant frequency band locates in the working regions of LISA \cite{Folkner:1998eni,amaroseoane2017laserinterferometerspaceantenna,LISA:2022yao,LISA:2022kgy,colpi2024lisadefinitionstudyreport}, Taiji \cite{TaijiScientific:2021qgx}, TianQin \cite{TianQin:2015yph,TianQin:2020hid}, DECIGO \cite{Kawamura:2023fwf,kawamura2020currentstatusspacegravitational}, ET \cite{abac2025scienceeinsteintelescope}, LIGO \cite{LIGOScientific:2007fwp,Harry:2010zz,LIGOScientific:2014pky,LIGOScientific:2016jlg,LIGOScientific:2018czr,LIGOScientific:2019lzm,LIGOScientific:2019vic,KAGRA:2021kbb}, VIRGO \cite{Cavalier:2000ecd,Chassande-Mottin:2010jdq,Marka:2010zz,Leroy:2013idg,Acernese:2015gua,Christensen:2017cte}, KAGRA \cite{KAGRA:2013rdx,KAGRA:2023pio}.

\section{Summary}\label{s:Summary}

We have shown that that both the initial PBH abundance and the present-day relic number density can be directly inferred from the location and amplitude of the peak of the IGW spectrum, provided that PBHs temporarily dominate the Universe before they evaporate. In this sense, the IGW spectrum acts as a cosmic diagnostic: The spectral peak point encodes the key PBH parameters and the iso-curvature fluctuation inherited from their Poisson statistics.
Remarkably, we find that the relevant frequency range is broad, making the scenario testable by multiple next-generation gravitational-wave observatories. 

In addition to gravitational waves, there are many ongoing or future projects to constain the physics of PBHs; 
microlensing surveys \cite{MACHO:2000qbb,EROS-2:2006ryy,Niikura:2017zjd,Niikura:2019kqi,li2025microlensingdarkmattermodels}, CMB spectral-distortion and anisotropy measurements \cite{Clesse:2014pna,Chluba:2019nxa,Lucca:2019rxf,Tashiro:2021kcg,CMB-S4:2023zem,Cyr:2023pgw,Tagliazucchi:2023dai}, BBN element-abundance constraints \cite{Iocco:2008va,Kawasaki:2008qe,Pospelov:2010hj,abazajian2012lightsterileneutrinoswhite,Baer:2014eja,Kawasaki:2017bqm}, and high-resolution gamma-ray and X-ray telescopes  \cite{Laha:2020ivk, Ghosh:2021gfa,Chianese:2025wrk,ouseph2025probingdoublepeakedgammarayspectra,Aljazaeri:2025ftv,50collaboration2025fermionicsubgevdarkmatter}.
Together with these, we will have a more solid pathway toward identifying or ruling out the PBH relic scenario with unprecedented precision. 

Our present analysis assumed a monochromatic PBH mass spectrum, no clustering, no hierarchical merging, and no accretion-driven mass evolution. 
These simplifying assumptions made us possible to isolate the cleanest IGW signatures from the PBH iso-curvature perturbation.
However, this idealized setup also means that any deviation between observations and our predictions encodes valuable information about the true PBH population and evolution. 
For instance:
Non-monochromatic mass functions will modify the IGW spectrum \cite{Bugaev:2010bb,Carr:2017jsz,Bellomo:2017zsr,Kohri:2018qtx,Inomata:2018epa,Pi:2020otn,Papanikolaou:2022chm,Kohri:2024qpd,bhaumik2025extendedmassdistributionpbhs}. 
Clustering may enhance the small-scale density contrast, leading to a larger IGW amplitude \cite{Belotsky:2018wph,Desjacques:2018wuu,Ding:2019tjk,Young:2019gfc,Domenech:2020ssp,LISACosmologyWorkingGroup:2023njw,He:2024luf,Auclair:2024jwj,Papanikolaou:2024kjb,Dizon:2025siw}.
Mergers and accretion may modify the effective evaporation time, shifting the IGW peak \cite{Nayak:2009wk,Raidal:2017mfl,Rice:2017avg,Cheek:2022mmy,Tabasi:2022fap,Thoss:2024hsr,Kohri:2024qpd,kalita2025revisitingpbhaccretionevaporation,aljaf2025mergerrateprimordialblack,Dienes:2025qdw}. 
We leave all of these for future works.

Therefore, precise measurements of the IGW spectrum—especially its peak position, width, and tail behavior—can serve as a powerful tool for reconstructing PBH mass functions, probing their clustering and merger histories, and testing early-Universe physics such as inflationary models \cite{Kawasaki:1997ju,Motohashi:2017kbs,Garcia-Bellido:2017mdw,Geller:2022nkr}, phase-transition \cite{Jedamzik:1996mr,Lewicki:2024ghw,Lewicki:2023ioy}, and topological defects \cite{Vilenkin:1982ks,Polnarev:1988dh,Matsuda:2005ey,James-Turner:2019ssu,jenkins2020primordialblackholescusp,SanglasMartin:2023ypg,Dunsky:2024zdo,Ferreira:2024eru}. In this way, a future detection (or non-detection) provides rich information far beyond the mere presence of PBH relics: it opens a window into the microphysics of the early cosmos.

\vskip 5pt

\paragraph*{Acknowledgements}

This work was supported in part by JSPS KAKENHI No.~24K00624. J.W. is also supported by Kavli IPMU, which was established by the World Premier International Research Center
Initiative (WPI), MEXT, Japan.



~\\~\\~\\~\\~\\~\\~\\~\\~\\~\\~\\~\\~\\~\\~\\~\\

\bibliography{Refs}

\end{document}